# Data Quality Certification using ISO/IEC 25012: Industrial Experiences

Fernando Gualo[1] [0000-0002-7800-7902],
Moisés Rodriguez[2] [0000-0003-2155-7409], Javier Verdugo [2] [0000-0002-2526-2918], Ismael Caballero[3] [0000-0002-5189-1427], Mario Piattini[3] [0000-0002-7212-8279]

[1] DQTeam, Paseo de Moledores s/n, 13071, Ciudad Real (Spain)
[2] AQCLab, Paseo de Moledores s/n, 13071, Ciudad Real (Spain)
[3] Alarcos Research Group, University of Castilla La Mancha, Institute of Technologies and Information Systems, Paseo de Moledores s/n, 13071, Ciudad Real (Spain)

fgualo@dqteam.es, mrodriguez@aqclab.es, jverdugo@aqclab.es, Ismael.Caballero@uclm.es, Mario.Piattini@uclm.es

**Abstract.** The most successful organizations in the world are data-driven businesses. Data is at the core of the business of many organizations as one of the most important assets, since the decisions they make cannot be better than the data on which they are based. Due to this reason, organizations need to be able to trust their data. One important activity that helps to achieve data reliability is the evaluation and certification of the quality level of organizational data repositories. This paper describes the results of the application of a data quality evaluation and certification process to the repositories of three European organizations belonging to different sectors. We present findings from the point of view of both the data quality evaluation team and the organizations that underwent the evaluation process. In this respect, several benefits have been explicitly recognised by the involved organizations after achieving the data quality certification for their repositories (*e.g., long-term organizational sustainability, better internal knowledge of data, and a more efficient management of data quality*). As a result of this experience, we have also identified a set of best practices aimed to enhance the data quality evaluation process.

## 1. Introduction

Regardless of their sector and across all geographies, organizations that are or intend to be data-driven use data for their operational, tactical, reporting, and strategic activities. Consequently, along with people, data can actually be considered as one of the most important assets for organizations [1][2]. This perception of data as an asset involves that organizations must commit firmly to the idea of data quality since the better their data is, the larger are the benefits they can obtain from using it [3][4]: it can be stated that data with adequate levels of quality can enable new ways to innovate in business in an increasingly competitive market [5]. Olson, in [6], and more recently, Redman in [7] demonstrated the connection between poor data quality – Redman reported losses around 3,1 billions of USD dollars in American companies due to poor data quality-, and the increase in cost and complexity in the development of systems [8]. In a similar vein, Friedman [9], Redman [10][11] and Laney [12] found that there are strong relationships between low data quality and some organizational concerns:

- Poor data quality is a primary reason why 40% of all business initiatives fail to achieve their target profit.
- Data quality affects overall labour productivity by as much as 20%.
- As more business processes become automated, data quality shows itself to be the limiting factor for overall process quality.

Bearing in mind all that has been pointed out above, it is clear that organizations should invest adequate resources in the deployment of mechanisms that help to ensure the reliability of the data; that is, data that has the appropriate level of quality for its present and future use [13]. Establishing these mechanisms to ensure and control the quality of data is a task that must be planned with sufficient time in advance, and it needs to be carried out with clear objectives and in conformity with the organization's strategy, so that suitably-skilled human resources, along with material and financial resources are allocated [14][15]. This is the only way to guarantee results which are proportional to the potential capability of the organization to exploit its data, either for internal purposes (e.g., to optimize specific aspects of the organizations) or for external purposes (e.g., to better serve their customers and partners).

In the light of our experience in the field of industrial certification of software quality [16], we propose that, in an analogous manner, certifying the level of quality of specific data repositories can provide customers and partners with the required confidence in that data. In [17], the environment established for data quality evaluation is presented. This environment consists of three main elements:

- A data quality model based on ISO/IEC 25012 [18] and ISO/IEC 25024 [19], which includes the data quality characteristics, as well as the underlying properties, measures and thresholds (see sections 2.1).
- A data quality evaluation and certification process based on ISO/IEC 25040 [20] which defines the activities, tasks, inputs, outputs and roles necessary to carry out a data quality evaluation (see section 2.2).
- A semi-automatic environment for data quality evaluation based on query tools and in-house software that generate evaluation results in accordance with the defined model and measures.

This data quality evaluation environment was developed by AQCLab[1], the first accredited laboratory by ENAC[2]/ILAC[3] for the evaluation of software product and data quality based on ISO/IEC 25000 standards. The evaluation reports resulting from the data quality evaluations performed on data repositories with this environment are the main input to the quality certification process carried out by AENOR[4] (the leading Spanish ICT certification body in Spain and Latin America).

This article presents three industrial experiences on the application of this environment in evaluation of data quality in three European organizations, as well as the findings we discovered from these experiences. This paper is structured as follows: Section 2 presents the basic concepts of the environment for data quality evaluations: the data quality model, the data quality evaluation process, and the way to measure data quality characteristics. Section 3 details how the three experiences of application of the environment in the industry were conducted and the results that were obtained. Section 4 presents further discussion regarding the findings from the application of the data quality evaluation process. Section 5 contains the threats to validity for the three industrial experiences. Finally, in section 6, we provide our conclusions on the evaluation and certification experience.

## 2. Data quality evaluation and certification using ISO/IEC 25000

In this section we present firstly the ISO/IEC 25000 (SQuaRE – Software product Quality Requirements and Evaluation) series of international standards intended for the evaluation and management of the quality of the components of information systems (see section 2.1).

Secondly, we provide an overview of the data quality evaluation and certification environment (see section 2.2), which has been used in the three industrial experiences presented in this paper. In addition, the data quality certification process is summarized in section 2.3.

### 2.1. ISO/IEC / 25000 – Data quality model and measures

The standards in the ISO/IEC 25000 series are organized in several divisions: quality management, quality models, quality measures and quality evaluation. As regards data quality evaluation, the most relevant standards in these divisions are ISO/IEC 25012 and ISO/IEC 25024.

ISO/IEC 25012 [18] identifies and provides a classification for the characteristics that define data quality. According to the standard, a data quality characteristic is "*the*

---

*category of data quality attributes that bears on data quality*". Regarding the classification of data quality characteristics, this standard categorizes them from two points of view:

— Inherent data quality, referring to the degree to which data quality characteristics have intrinsic potential to satisfy implicit data needs.
— System-dependent data quality, referring to the degree to which data quality is achieved and preserved through an information system and is dependent on the specific technological context in which the data is used.

Due to the wide range of differences in the technological nature of data repositories, which makes the development of generic measures that would allow the comparison between different organizations almost impossible, only the inherent data quality characteristics have been considered for the data quality evaluation environment. This way, the implementation of the subsequent measures used in the evaluation can be taken intrinsically and independently of the specifities of the dataset and the technology of the information system that supports the data repository. Thus, the process of generating data quality measures is repeatable for any dataset in any domain, and the results can be compared and benchmarked. The inherent data quality characteristics are described in **Table 1**.

| Characteristic | Definition |
|---|---|
| *"Accuracy"* | The degree to which the data has attributes that correctly represent the true value of the intended attribute of a concept or event in a specific context of use. |
| *"Completeness"* | The degree to which subject data associated with an entity has values for all expected attributes and related entity instances in a specific context of use. |
| *"Consistency"* | The degree to which data has attributes that are free from contradiction and are coherent with other data in a specific context of use. |
| *"Credibility"* | The degree to which data has attributes that are regarded as true and believable by users in a specific context of use. |
| *"Currentness"* | The degree to which data has attributes that are of the right age in a specific context of use. |

**Table 1.** Inherent data quality characteristics defined in ISO/IEC 25012 [18].

ISO/IEC 25024 [19] is the part of the SQuaRE series of international standards that establishes the relationship between the concept of data quality characteristic (introduced in ISO/IEC 25012) and the concept of "quality property". A quality property is an element that represents a way of evaluating certain aspects or particularities of the data contained in a repository.

To evaluate the quality of a data repository, an organization should identify the data quality characteristics and the corresponding data quality properties that best fit their stated data quality requirements. **Fig. 1** shows a summary of the inherent data quality characteristics and the data quality properties defined for each of them.

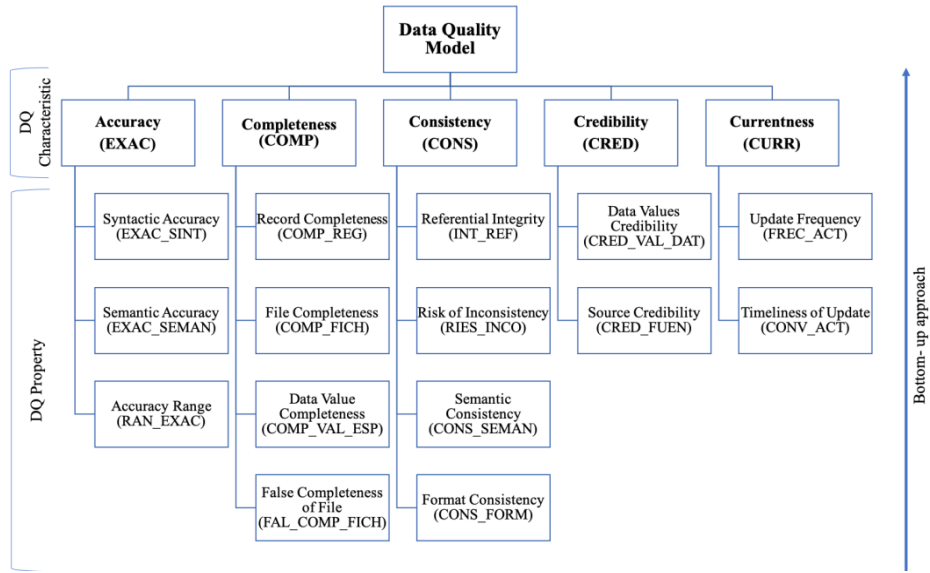

**Fig. 1.** Inherent data quality characteristics and related data quality properties extracted from ISO/IEC 25012 and ISO/IEC 25024, respectively.

**Table 2** shows an example of how each data quality property is described in ISO/IEC 25024, and the information it provides on how to calculate their value. The evaluation team must interpret when low values for the measurement of a property represent an issue in the data repository.

| Data quality characteristic | Accuracy |
|---|---|
| Data quality property | Data Accuracy Range |
| Measurement description | Data Accuracy Range focuses on checking if data values are included in the required intervals. Its value is obtained as the ratio of records in a data file whose values for their fields are within the specified intervals. |
| Calculation formula | X=A/B<br>A= number of data items having a value included in a specified interval (i.e., range from minimum to maximum)<br>B= number of data items for which can be defined a required interval of values |
| Scale | Ratio |
| Value range | [0.0, 1.0] |

**Table 2.** Description for the property *"Accuracy Range" (RAN_EXAC)* and its measurement.

### 2.2. Data quality evaluation environment

ISO/IEC 25024 does not specifically address how the measures corresponding to the data quality properties should be aggregated to compute the overall quality level for each data quality characteristic. It is a matter for the organization conducting the data quality evaluation to adequately specify how to aggregate the measures for the data quality properties in order to obtain the quality level value for the data quality characteristics, as well as to establish the specific data quality evaluation process.

The evaluation process - which comprises the activities, inputs, and outputs required to conduct a data quality evaluation (see **Fig. 2**) – carried out by personnel from AQCLab – the evaluation team – consists of five systematic, rigorous and repeatable activities through which to obtain the measures for the selected data quality characteristics. The most relevant aspects for each activity of the accredited laboratory's data quality evaluation process are described below.

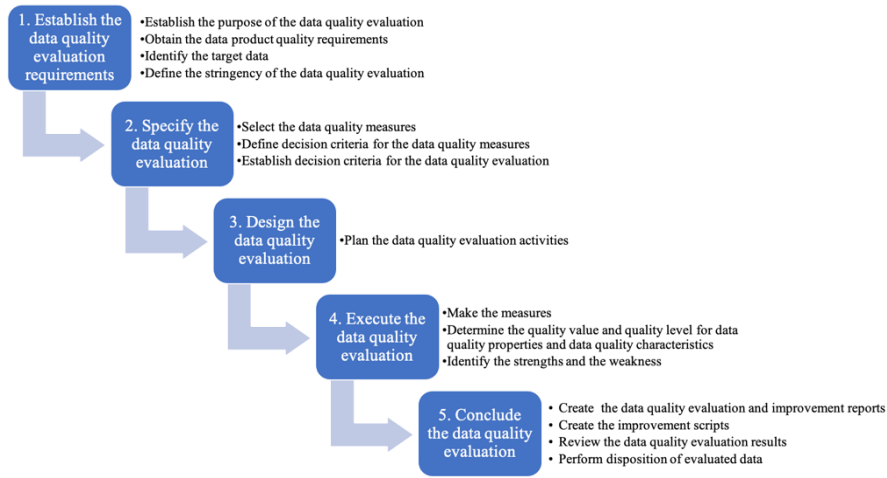

**Fig. 2.** Data quality evaluation process established by the accredited laboratory (adapted from [20]).

During the **Activity 1: Establish the data quality evaluation requirements**, several meetings and working sessions are held between the evaluation team of the accredited laboratory and the staff of the organization seeking to evaluate the data quality level of a given repository. This staff for the organization under evaluation must include stakeholders who know the business at a higher, managerial level, and personnel who are more involved with the technical and operational side. During these sessions, the data quality model and measures are introduced by the evaluation team to the staff of the organization with the aim of selecting those data quality characteristics that best fit their data quality requirements, and consequently, they want to have evaluated. In this regard, organizations can select all or only a subset of the data quality characteristic defined in the data quality model. Besides, during this activity, the organization staff pre-sents the data repository to be evaluated to the evaluation team, along with the documentation where its business rules are defined. Based on the above, the scope of the evaluation is outlined in this activity, being subject to progressive refinement during activities 2 and 3 of the process. The scope of the evaluation takes into account the needs of the organization regarding the purpose of their data quality management initiatives, and it is determined by two factors: (1) the number of data entities (e.g. tables and views) that are part of the data repository to be evaluated, and (2) the number of business rules that each element of the data repository (e.g. data attribute for a relational table) must comply with in relation to the data quality characteristics and data quality properties to be evaluated. Typically, the purposes for the data quality management initiatives of the organization as regards the data quality evaluation include detecting data defects on the data repository, assuring the quality of the data repository, certifying the data repository in terms of its quality, or comparing the quality of the data repository with that of the competition. From the point of view of the accredited laboratory, it is of paramount importance that the organizations first define the business rules for their data repositories with a sufficient degree of detail, since these business rules are the basis for the evaluation process. This way, the document containing all the business rules for the data repository must be reviewed by the stakeholders and the evaluation team until an agreement on its adequacy is reached. After that, the staff of the organization with the support of the evaluation team must categorize the business rules into the data quality properties of the evaluation model, considering the nature of the business rules and the intent of the data quality properties.

Furthermore, during this activity, and in order to shape the data quality evaluation plan, some other points are discussed and agreed upon, such as the way to access the data repository clone and the necessary security constraints (e.g., VPN connection), the time restrictions for the connection so as to not interfere with the operation activities of the organization, as well as the deadline for the provision of the results of the evaluation.

As a result of this activity, the main inputs for the evaluation should be identified and made available, being these main inputs the following:

- Static clone of the data repository.
- Document specifying the business rules for the data repository (see examples in Table 3), which has been agreed upon by the organization and the evaluation team, and in which the data requirements that the data repository must comply with are gathered.
- The set of selected data quality characteristics to be evaluated.

| Business rule ID | Business rule example |
|---|---|
| 1 | person.id must be represented with 9 characters, in which the first eight must be digits in the range [0-9], and the last one must be an alphabetic character in the range [A-Z]. |
| 2 | person.ipaddress must be expressed as four numbers in the range [0, 255] separated by a period, such as 126.12.4.89. |
| 3 | warning.type must take one of the following values: {IT GENERAL, SUPERCOMPUTATION, HR}. |

**Table 3.** Examples of business rules for a data repository.

**Activity 2: Specify the data quality evaluation.** Once the data quality characteristics have been introduced and selected in the Activity 1, and with them, the corresponding data quality properties (see **Fig. 1**), in this activity the measurement methods and the corresponding quality levels for the measures of each data quality property are also presented, so that the organization can understand the results of the evaluation. This understanding will enable the refinement of the information provided about the data repository as well as the tuning of some implementation details of the business rules for the data repository. This knowledge will be used by evaluation team when carrying out the Activity 4 in order to develop the evaluation scripts.

The quality levels and measurement methods (as well as the underlying elements, like the threshold values used in the corresponding measurements) have been defined by the accredited laboratory as part of the tailoring of the data quality model, which was supported by several pilot experiences. For instance, the threshold values used in the calculation of the data quality levels have been tuned to represent how well organization can perform at different scenarios, where inadequate levels of data quality are the source of different issues. The higher the quality level is (and the corresponding threshold value), the less potential impact the data quality issue can have on the organization.

As a result of this activity, the following output is obtained:
- Understanding of the specific details of the data quality evaluation process by the organization.

During the **Activity 3: Design the data quality evaluation**, the scope of the evaluation is refined and completely defined in detail. In addition, the evaluation plan containing the activities and tasks needed to conduct the evaluation is completely established.

The following product is obtained as a result of this activity:
- Data quality evaluation plan, containing the specific details (e.g., schedule, tasks, resources, etc.) for conducting the data quality evaluation over the data entities included in the scope.

With the **Activity 4: Execute the data quality evaluation**, the quality level for the selected data quality characteristics are obtained by following the evaluation plan defined during activities 1 to 3. These quality level values for the quality characteristics are obtained following a hierarchical bottom-up approach as shown in **Fig. 3**.

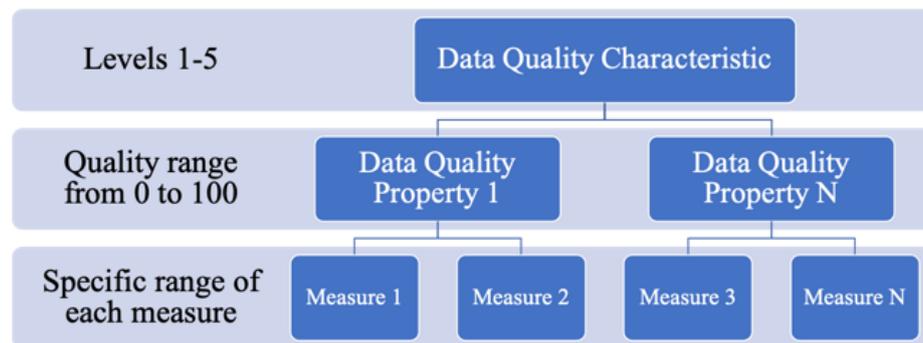

**Fig. 3.** Hierarchy of the elements that comprise the data quality model for the evaluation of data quality repositories.

This activity involves the use of some concepts (e.g., quality levels, quality ranges and profiling functions) that are explained in detail in [16]. The tasks required in the execution activity are the following:

1) ***Design the evaluation scripts.*** Taking as input the business rules already categorized for each data quality property and quality characteristic, as well as any other documentation provided by the organization seeking evaluation and the information obtained in the meetings during the **Activity 1**, the evaluation team develops a set of evaluation scripts that will allow to check each business rule defined for each entity in the scope of the evaluation. **Code Listing 1** shows an example of an evaluation script that is aimed to check the compliance of the business rule 1 from **Table 3**, categorized under the *"Syntactic Accuracy"* quality property.

```
SELECT COUNT(*) FROM exampleDB.person p1
WHERE p1.id IS NOT NULL AND p1.id REGEXP '^[0-9]{8}[A-Za-z]{1}$'
UNION
SELECT COUNT(*) FROM exampleDB.person p2;
```

**Code Listing 1.** Example of evaluation script implemented in SQL language (MySQL) for the validation of business rule ID 1 in Table 3.

2) ***Execution of the evaluation scripts.*** Following the evaluation plan, the evaluation scripts for each entity are executed in order to obtain the base measures for each data quality property, according to what is defined in ISO/IEC 25024 (see **Table 2**). Based on the results of this execution, the weaknesses and strengths for each data quality property can be outlined. It is important to remark that even if the definition of the measure method is done according to ISO/IEC 25024, applying it requires customization as regards the specific semantics of the data entities under evaluation and the technological aspects of the system of which the data repository under evaluation is part.

3) ***Production of the quality value for the data quality properties.*** The base measures obtained previously are used to calculate the quality value for each data quality property by applying over them evaluation functions. It must be taken into account that these evaluation functions are under industrial property rights protection, and because of that, no further details can be reproduced in this paper. The quality values for quality properties provided by said functions are in the range [0, 100]. At this point, the weaknesses and strengths previously identified for each data quality property should be analysed and discussed, in order to identify the critical aspects that are susceptible to be improved.

4) ***Derivation of the quality level for the data quality properties from their quality value.*** The evaluation for the data quality characteristics is performed by using a specific profiling function that requires to take an intermediate step in which the quality values for each data quality property are transformed into quality levels. This transformation is performed by taking a set of ranges delimited by threshold values, as the ones shown in **Table 4** as an example, and classifying the quality value accordingly. Once again, the specific values for the thresholds used by the laboratory are under industrial property rights protection, and they cannot be reproduced here.

| Level | Quality value |
|:-----:|:-------------:|
| 1 | 0 – 20 |
| 2 | 20 – 40 |
| 3 | 40 – 70 |
| 4 | 70 – 85 |
| 5 | 85 – 100 |

**Table 4.** Example of quality value ranges used in the determination of quality levels for data quality properties.

5) ***Determination of the quality level for the selected data quality characteristics***. In this step, the quality level for each data quality characteristic is calculated by aggregating the quality levels of the corresponding data quality properties (see **Fig. 1**). This aggregation is performed each characteristic by means of a profiling function (see [16]), such as the one shown in **Table 5**. These functions use the concept of profiles, which are vectors that represent the amount of corresponding quality properties in each quality level for a quality characteristic. A profiling function defines a set of ranges, which specify the maximum number of data quality properties admissible in each quality level in order to determine if the quality characteristic is in that range. Then, the profile of the system for a characteristic (the amount of quality properties in each quality level) is compared against those ranges, which allows to determine the quality level reached for the quality characteristic. As an explanatory example of how this profiling function works, the following shows how it could be applied to the "*Accuracy*" data quality characteristic, which is evaluated through three data quality properties (namely, "*Syntactic Accuracy*", "*Semantic Accuracy*", and "*Accuracy Range*"). Let us suppose that in the previous step 4, we have obtained the following quality levels for the properties: 4 for "*Syntactic Accuracy*", 4 for "*Semantic Accuracy*", and 3 for "*Accuracy Range*"). These values give the profile <0, 0, 1, 2, 0> for this characteristic, which means that there are no data quality properties at levels 1 and 2, there is one property at level 3, two at level 4, and none at level 5. When comparing this profile to the ranges defined in Table 5 we see that range 5 allows no properties with a quality level lower than 5, and range 4 does not allow properties with a quality level lower than 4. Since the profile the characteristic has one characteristic at level 3, these ranges cannot be met. On the other hand, range 3 allows a maximum of 2 data quality properties at level 3, one at level 2 and none at level 1. The profile meets the requirements for this range, and, consequently, in can be determined that the quality level value for the *"Accuracy"* characteristic in this example is 3 for the repository under study. This step is performed in a similar vein for each of the selected data quality characteristics, applying its own profiling function. One important concern about the profiling functions is that the way in which they are built, they provide a fair aggregation of values, preventing that this aggregation leads to misleading data quality levels, as it could happen with the use of, for example, the mean value of the quality properties. Since the specific profiling functions used by the laboratory are under industrial property rights protection, they cannot be reproduced in this paper.

| | | Levels | | | | Data quality characteristic value |
|---|---|---|---|---|---|---|
| | | 1 | 2 | 3 | 4 | |
| **Range** | **0** | - | - | - | - | 0 |
| | 1 | 3 | 3 | 3 | 3 | 1 |
| | 2 | 2 | 3 | 3 | 3 | 2 |
| | 3 | 0 | 1 | 2 | 3 | 3 |
| | 4 | 0 | 0 | 0 | 3 | 4 |
| | 5 | 0 | 0 | 0 | 0 | 5 |

**Table 5.** Profiling function proposed as an example to evaluate the "*Accuracy*" data quality characteristic (example adapted from [16]).

As a result of this activity, the following outputs are obtained:
- Quality value and quality level for each data quality property evaluated.
- Quality value for each data quality characteristic selected for the evaluation.
- Weaknesses and strengths detected as regards each data quality property.

Finally, in **Activity 5: Conclude the data quality evaluation** a detailed evaluation report is produced, reflecting the quality levels achieved for the selected data quality characteristics, as well as the values obtained for the corresponding quality properties. Additionally, a comprehensive improvement report can be produced and provided to the organization, detailing the weakness and strengths related to each data quality property. This improvement report focuses on those properties that did not achieve an adequate level of quality, and provides detail on the causes for such low levels, so that the organization can take action to improve them. In addition, a set of scripts are provided to the organization identify the specific records requiring improvement actions. Lastly, the access provided to the data repository and the other assets necessary for the data quality evaluation is removed or revoked.

As a result of this activity, the following work products are generated:
- Evaluation report, detailing the results of the evaluation.
- Improvement report, if necessary.
- Report informing the revocation of the access privileges on the evaluated data repository.
- Feedback from the organization regarding their perception of the data quality evaluation process, reviewed by the evaluation team in order to continually improve it.
- Scripts related to the improvement report, facilitating the identification of the data records that did not meet adequate levels of quality (weaknesses), thus requiring improvement actions from the organization.

### 2.3. Data quality certification

An organization that is interested in certifying the quality level of a data repository can request the start of the data quality certification process once the data repository has been evaluated. To be eligible for a data quality certificate from a certification body, all the data quality characteristics evaluated must reach a specific quality level, determined by the certification body. AENOR, the certification body associated with the laboratory, has established that each data quality characteristic evaluated must reach at least the quality level 3 in order to be able to grant a certification for the requesting organization.

The data quality evaluation and certification process is similar to the one presented by Rodriguez et al. in [21], and it consists of the following steps (see **Fig. 4.**):

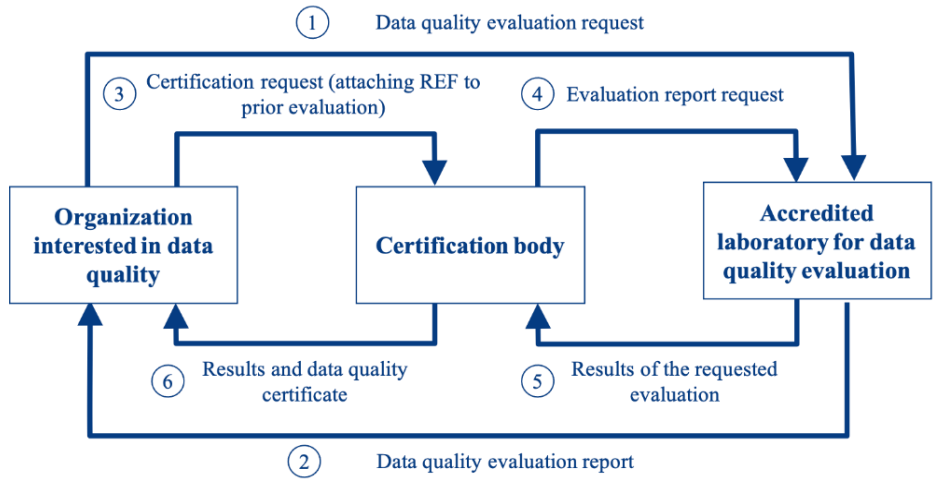

**Fig. 4.** Data quality certification process (adapted from [21]).

Firstly, the organization interested in certifying the data quality levels for a repository contacts an accredited laboratory (such as AQCLab), and after establishing the scope of the data quality evaluation, they sign a contract stating the terms of the collaboration (e.g., the deliverables of the evaluation, phases of the process, as well as some other concerns like data privacy responsibilities and accountability). Once the contract is signed by both parties, the evaluation process begins and is carried out by following the activities described in section 2.2. After the finalization of the evaluation, the accredited laboratory issues the evaluation report detailing the results regarding the data quality results of the evaluated characteristics. Depending on these results, the organization must decide whether they want to improve the quality of the data repository or start the certification process if they have reached the required quality level values. When the organization consider that they are ready to get their data repository certified, they must contact the certification body to request the data quality certification, providing them with the reference of their evaluation report. After that, the certification body contacts the accredited laboratory that carried out the evaluation in order to verify the results of the report referenced by the organization; the accredited laboratory provides the certification body with the information requested. After verifying the validity of the results of the evaluation, the certification body conducts a confirmation audit on the organization and their repository, and finally issues and grants the organization with the corresponding data quality certificate. This last step carried out by the certification body has a pre-established cost and a duration of two days, independently of the characteristics of the data repository to be certified.

## 3. Industrial experiences

This section describes the experience about the application of the data quality evaluation and certification environment in three industrial experiences.

Each of the three industrial experiences presented in this section has followed the phases presented in **Fig. 5**.

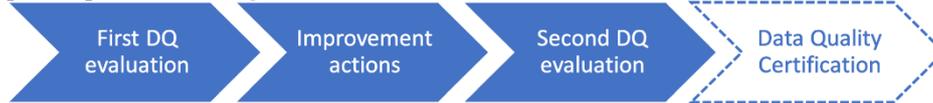

**Fig. 5.** Phases carried out in each data quality evaluation in the three industrial experiences (the specific activities for each evaluation are shown in **Fig. 2**).

For each of the three industrial experiences, a first complete data quality evaluation was initially carried out by the accredited laboratory by following the evaluation process specified in **Fig. 2**; secondly, due to the results obtained in that first evaluation, improvement actions were carried out by the organizations on their data repositories in order to address the weaknesses detected. After that, a second complete data quality evaluation was carried out by the accredited laboratory (also following the evaluation process specified in **Fig. 2**). Finally, after this second evaluation, the organizations reached the necessary levels to certify the quality of their data repositories. After contacting the certification body, they were granted with the certificate of data quality for their repositories.

### 3.1. Organizations involved in the industrial experiences

Prior to describing the organizations involved in the three industrial experiences, we want to remark two important matters in this section. Firstly, due to confidentiality agreements, we are not allowed to explicitly name the organizations under evaluation, so instead, we will refer to them only by means of the sector to which they belong. Secondly, since there are still not many cases of organizations that have obtained a data quality certification, we must keep the particular evaluation results anonymous. Because of these two reasons, the organisations involved will be identified as Org.1, Org.2 and Org.3; and the order in which the industrial experiences are detailed in the following subsections does not correspond to the order in which the description of the organizations is presented in the following paragraph.

One of the organizations is a travel agency with more than 1200 employees, whose purpose is to sell different types of trips, tours and combinations, ranging from flights and train tickets to hotel stays and vacation packages. The main purpose for this organization to seek data quality evaluation and certification was to ensure that their data was good enough to use as input in the creation of several BI applications (business dashboards). Another organization, with 400 employees, is a business school that offers different types of higher education programs, masters, and courses, as well as support for organizations in the development of their digital talent initiatives. The motivation of this organization for seeking data quality evaluation was to determine how good the data gathered through teaching quality surveys (answered by students after the finalization of their programs) was; this data was intended to be used to improve their course offering. The third organization is a business registry with more than 2000 employees; its purpose is to keep track of the existing business organizations and all the activities they carry out. Their intent with the evaluation was to obtain an overview of the quality of the data contained in their repository and to ensure appropriate levels of quality for that data.

### 3.2. Experience 1. Organization 1 (Org. 1)

The scope of the data quality evaluation for Org.1 was discussed and established during several initial meetings with different staff of the organization. The evaluation was to be carried out on the repository that contains relevant data for conducting their core business, as well as some other operations. An overview of the scope of the repository evaluated is shown in **Table 6**.

| **Database technology** | | SQLServer 2017 (version 14.0) |
|---|---|---|
| **Number of entities** | | 14 tables |
| **Number of records** | | 10 million |
| **Data quality characteristics evaluated** | | *"Accuracy", "Completeness", "Consistency", "Credibility", and "Currentness"* |
| **Business rules involved in the evaluation** | *"Accuracy"* | 89 |
| | *"Completeness"* | 78 |
| | *"Consistency"* | 91 |
| | *"Credibility"* | 54 |
| | *"Currentness"* | 63 |
| **Total business rules involved in the evaluation** | | 375 |

**Table 6.** Summary of the scope for the evaluation of Org. 1's data repository.

A clone of the repository was created and deployed in a server by the organization; access to the clone was provided to the evaluation team; and the corresponding evaluation scripts were created. In order to carry out the first data quality evaluation in Org. 1, and considering the characteristics of the data repository, as well as the scope and the number of business rules, a team of two evaluators was assigned by the accredited laboratory. The team invested the following effort in the different activities of the evaluation:

- 60 days in collaboration with Org. 1 for the identification of the business rules, the creation of the business rules document and the implementation of the evaluation scripts.
- 13 days to execute the evaluation scripts, to collect the results and to determine the data quality values for the quality properties, as well as the quality level for the properties and the data quality characteristics.

- 3 days to produce the evaluation report, the improvement report, and the improvement scripts to locate the weaknesses identified in the repository through the data quality evaluation.

After the first evaluation (see **Fig. 6**), the results showed that only the characteristic *"Credibility"* achieved the required quality level for certification (level 3 or higher). *"Accuracy"* and *"Consistency"* achieved level 1, while *"Consistency"* and *"Current-ness"* achieved level 2, which were not enough for their certification according to the certification body's criteria.

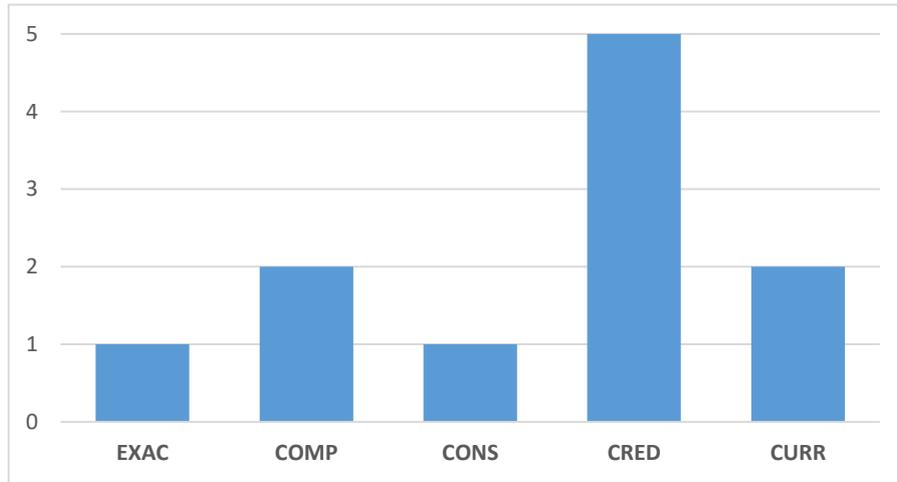

**Fig. 6.** Results of the first data quality evaluation for Org. 1.

**Table 7** shows an example of the strengths and weaknesses found during the first data quality evaluation for Org. 1.

| Data quality characteristic | Strengths | Weaknesses |
|---|---|---|
| **Accuracy** | | • The data values must match certain defined regular expressions.<br>• The data values must be defined within the acceptability ranges of the value domain defined for the attributes. |
| **Completeness** | • The data files contain an appropriate number of records to represent the required entities. | • For a full representation of an entity, all |
| | • Data attributes contain all the values required for the different entities represented by the data.<br>• There is no false file completeness, as existing records are adequate, complete and unduplicated. | the attributes necessary for the business must have a value. |
| **Consistency** | | • The data must comply with generically defined formats for equivalent data types. |
| **Credibility** | • The data is highly reliable, since it is entered and used by Org. 1 personnel as part of its business processes.<br>• Data values are highly reliable, since they go through validation processes. Additionally, surveys reveal that the data is considered highly credible by the users of the data themselves. | |
| **Currentness** | • Data that need to be updated under specific time periods are updated frequently in the Org. 1 data repository. | • Data must be updated for events that occur under certain conditions defined by the organization. |

**Table 7.** Example of strengths and weaknesses of the data of Org. 1 detected in the first evaluation.

Considering the strengths and weaknesses identified during the first evaluation, Org. 1 planned a meeting to discuss the details for the improvement process. In this meeting, Org. 1 agreed to address all the proposed improvements in order to raise the quality of their data and meet the quality level required (3 or higher) for the certification of all the data quality characteristics. Once these improvements were discussed, an improvement implementation plan was established by Org.1. The improvements were carried out in two ways. The first one focused on reviewing and adapting the business rules defined to produce an improved version of the business rules document. The second one focused on solving the identified weaknesses on the data included in the repository. Org. 1 assigned 3 people to implement the proposed data quality improvements: one person focused on reviewing and adapting the business rules document, while the other two reviewed the records in the repository, and made the necessary changes to remediate all the weaknesses indicated in **Table 7**. In order to facilitate the identification of the records to be modified, the accredited laboratory provided a set of scripts that determined which records did not comply with the business rules. Therefore, the improvement actions were carried out without the intervention of the accredited laboratory. It took Org. 1 8 weeks approximately to resolve all the weaknesses found as a result of the first data quality evaluation.

After that, a second data quality evaluation was carried out on a new version of the data repository. This second evaluation was carried out in 17 days: 3 days to review, modify and create the necessary evaluation scripts, 13 days to execute the evaluation, and 1 day to create the evaluation report.

Because of the improvements that were implemented by the organization, the value of the data quality properties increased considerably. A comparison of the results obtained for each of the data quality properties between the first and the second evaluation is presented in **Fig. 7**.

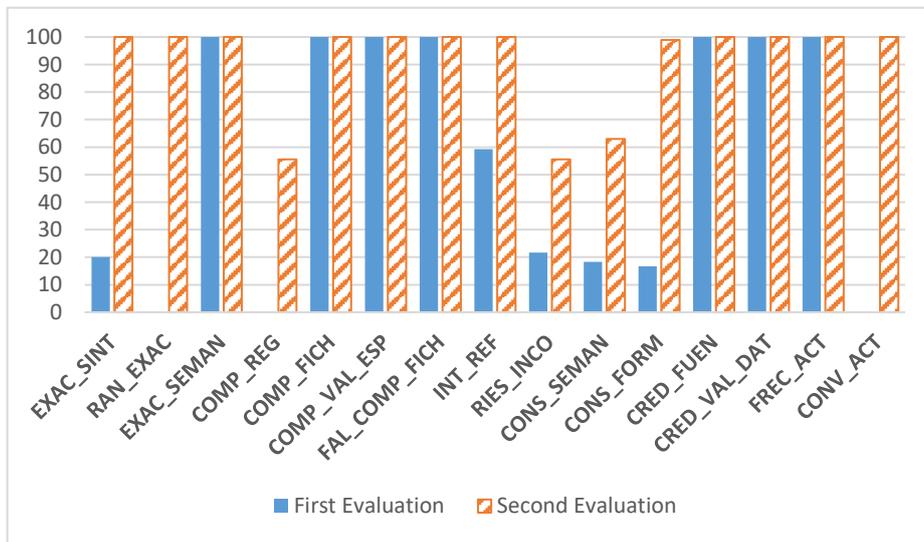

**Fig. 7.** Comparison of quality values for properties between the first and the second data quality evaluation for Org. 1.

Consequently, the quality level of the data quality characteristics was also improved. The conclusions of the second data quality evaluation are shown below:

- The "*Accuracy*" data quality characteristic was improved from level 1 to level 5 after solving the weaknesses related to the *"Syntactic Accuracy"* and *"Semantic Accuracy"* properties.
- The "*Completeness*" characteristic was improved from level 2 to level 4 after solving many weaknesses for the "*Record Completeness*" property.
- For the "*Consistency*" characteristic, the organization achieved level 3 after remediating weaknesses identified for all the corresponding data quality properties.
- Finally, Org. 1 defined new business rules for the data quality property *"Timeliness of Update"*, which allowed them to reach level 5 for the *"Currentness"* data quality characteristic.

### 3.3. Experience 2. Organization 2 (Org. 2)

Org. 2 decided to evaluate all the data quality characteristics for the data repository that contains information for their core activity (see **Table 8**). Analogously to the previous experience, the evaluation was carried out on a clone of the repository accessed by an ad-hoc VPN connection.

| Database technology | | Oracle 18c |
|---|---|---|
| Number of entities | | 36 |
| Number of records | | 750 million |
| Data quality characteristics evaluated | | *"Accuracy", "Completeness", "Consistency", "Credibility", and "Currentness"* |
| Business rules involved in the evaluation | *"Accuracy"* | 189 |
| | *"Completeness"* | 131 |
| | *"Consistency"* | 340 |
| | *"Credibility"* | 72 |
| | *"Currentness"* | 81 |
| Total business rules | | 813 |

**Table 8.** Summary of the scope of the evaluation for Org. 2's data repository.

Considering the characteristics of Org. 2's data repository, two people participated as part of the evaluation team, and the following effort (in days) was dedicated to each activity of the process for the first evaluation:

- 125 days in collaboration with Org. 2 for the identification of the business rules, the creation of the business rules document and the implementation of the evaluation scripts.
- 19 days to execute the evaluation scripts, to collect the results and to determine the quality values for data quality properties, as well as the quality levels for the properties and data quality characteristics.
- 3 days to create the evaluation report, the improvement report, and the improvement scripts to remediate the weaknesses identified during the data quality evaluation.

After the first evaluation (see **Fig. 8**), it was determined that the characteristics *"Completeness", "Credibility"* and *"Currentness"* achieved the necessary level to attain certification, while *"Accuracy"* and *"Consistency"* achieved level 1, thus not being eligible for certification.

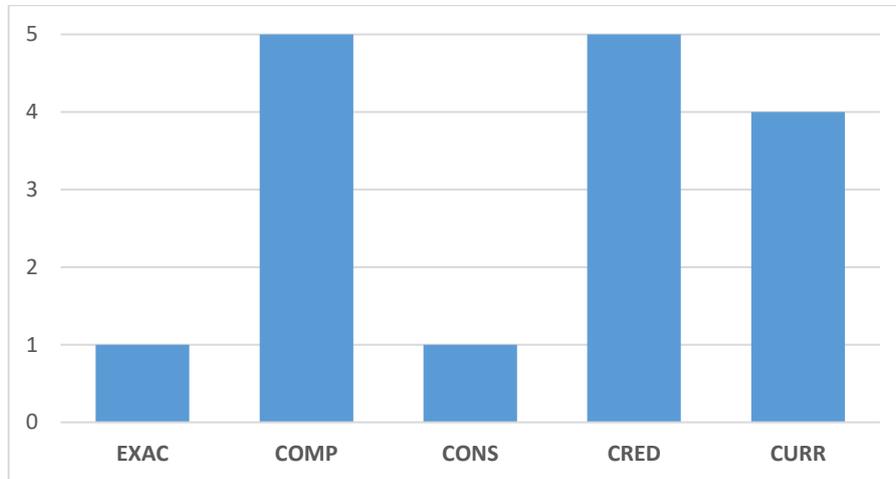

**Fig. 8.** Results of the first data quality evaluation of Org. 2.

Considering the strengths and weaknesses identified for the characteristics *"Accuracy"* and *"Consistency"* in the first evaluation, Org. 2 decided to address the weaknesses in order to improve their data and be able to certify all the data quality characteristics. Thus, Org. 2 started an improvement process to fix all the weaknesses identified. The accredited laboratory produced and provided the scripts to identify which records did not meet the business rules defined, and Org. 2 carried out the specific improvements independently. Org. 2 assigned a team of 8 people to implement the improvement actions, which took them around 3 weeks. The improvement team of Org.2 had a first planning meeting in which they reviewed the business rules documentation, identified the necessary changes and planned the implementation of the improvement actions. After that, the improvement team began to make the changes required so that the values of the records would comply with the business rules.

Once these improvements were implemented, the evaluation team reviewed the updated business rules documentation, reviewed and adapted the evaluation scripts, and carried out a second data quality evaluation on the new version of the data repository. Thanks to the previous knowledge of the evaluation team as regards the data repository, the second evaluation was conducted by 2 people in 23 days: it took them 3 days to modify and create the evaluation scripts, 19 days to execute the scripts and obtain the results of the evaluation and 1day to create the evaluation report. A comparison of the results obtained for the first and the second evaluation is presented in **Fig. 9**.

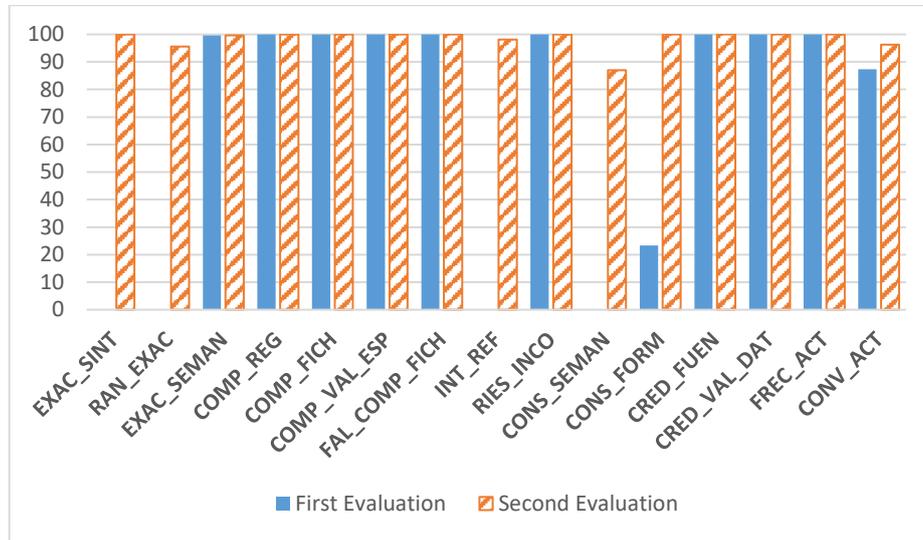

**Fig. 9.** Comparison of results for quality properties between the first and the second data quality evaluation for Org. 2.

During this second evaluation, the improvement efforts of Org. 2 focused on those data quality characteristics whose quality level could be improved: *"Accuracy", "Consistency", and "Currentness"*. The data quality characteristics *"Completeness"* and *"Credibility"*, which obtained quality level 5 in the first evaluation, were also evaluated in the second one, reaching the same quality level. The conclusions of the second evaluation were:

- The *"Accuracy"* data quality characteristic was improved from level 1 to level 5 after solving the weaknesses related to the data quality properties *"Syntactic Accuracy"* and *"Semantic Accuracy"*.
- The organization improved the *"Consistency"* characteristic from level 1 to level 3 by addressing weaknesses identified for the data quality properties *"Referential Integrity", "Semantic Consistency",* and *"Format Consistency"*.
- Finally, the quality characteristic *"Currentness"* reached level 5 after defining new business rules for the data quality property *"Timeliness of Update"*.

### 3.4. Experience 3. Organization 3 (Org. 3)

At the beginning of the evaluation project, Org. 3 selected all the data quality characteristics to be included in the scope of the evaluation of their data repository. This repository contained data from their operational activities that was used for making decisions at the managerial level. A full summary of the data repository is shown in **Table 9**.

| Database technology | | Teradata Database 15.0 |
|---|---|---|
| Number of entities | | 10 |
| Number of records | | 20 million |
| Data quality characteristics evaluated | | *"Accuracy", "Completeness", "Consistency", "Credibility", and "Currentness"* |
| Business rules involved in the evaluation | *"Accuracy"* | 94 |
| | *"Completeness"* | 100 |
| | *"Consistency"* | 176 |
| | *"Credibility"* | 48 |
| | *"Currentness"* | 70 |
| Total business rules | | 488 |

**Table 9.** Summary of the scope for the quality evaluation of Org. 3's data repository.

After discussing all the aspects regarding the scope of the evaluation, a clone of the repository was created and deployed, and access to it was granted to the evaluation team through a VPN connection. Considering the technological environment of the data repository, the scope of the evaluation, and the number of business rules, the accredited laboratory allocated people to the evaluation team. The evaluation took the following effort for the different activities of its process:

- 90 days in collaboration with Org. 2 for the identification and refinement of the business rules, the creation of the business rules document and the implementation of the evaluation scripts.
- 15 days to execute in a semi-automatic way the more than 450 evaluation scripts, collecting the results and determining the quality value for quality properties, and obtaining the quality level for properties and data quality characteristics.
- 3 days to create the evaluation report and the improvement report, and implement the improvement scripts to locate the weaknesses identified during the data quality evaluation in the specific records of the data repository.

Based on the business rules description provided by Org. 3, the evaluation scripts were created by the evaluation team to check their compliance against the data repository. After the first evaluation (see **Fig. 10**) only the data quality characteristic *"Accuracy"* did not achieve the quality level threshold value required for certification.

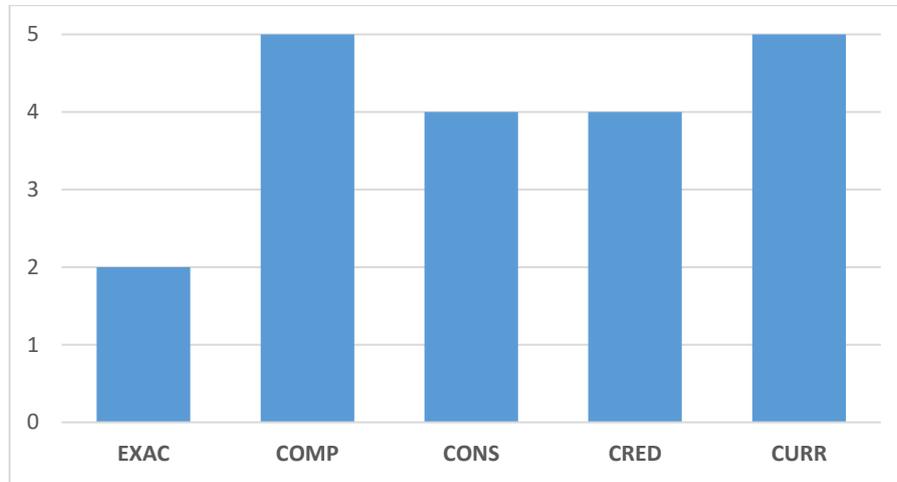

**Fig. 10.** Results of the first data quality evaluation of Org. 3.

Like Org.1 and Org. 2, the stakeholders of Org. 3 met to review the evaluation report, and the strengths and weaknesses identified during first evaluation. Finally, they decided to address only the improvements identified as necessary for the data quality property *"Accuracy Range"*, since they considered that the results obtained for the other data quality characteristics were good enough. In relation to this, Org. 3 assigned a team of 2 people, and it took them 3 weeks to carry out the improvement actions. In this case, the business rules documentation did not require any revision for updates, and they focused solely on remediating the weaknesses found on the data itself. As with previous cases, the improvement scripts provided by the accredited laboratory were used to identify the records that did not comply with the business rules defined and, thus, had to be corrected to improve the data quality level of the repository. Using these scripts, Org. 3 was able to address on their own the implementation of the necessary changes in the data repository. They solved the data quality problems by adding new controls and by modifying their data processing automations.

After the improvement process by Org.3, a second evaluation was conducted by the evaluation team. This second evaluation was carried out by 2 people in 18 days: 4 days to create and modify the evaluation scripts, 13 days to execute the evaluation scripts and process the results, and 1 day to create the evaluation report. The time required for the second evaluation, as usual, was less than that for the first evaluation, since the evaluation team can take advantage of the knowledge and the outputs generated during the first evaluation. After the second data quality evaluation, the value for the *"Accuracy Range"* property improved considerably, and enabled Org. 3 to reach quality level 5 for the characteristic *"Accuracy"*. A comparison between the results obtained in the first and the second evaluation is presented in **Fig. 11**.

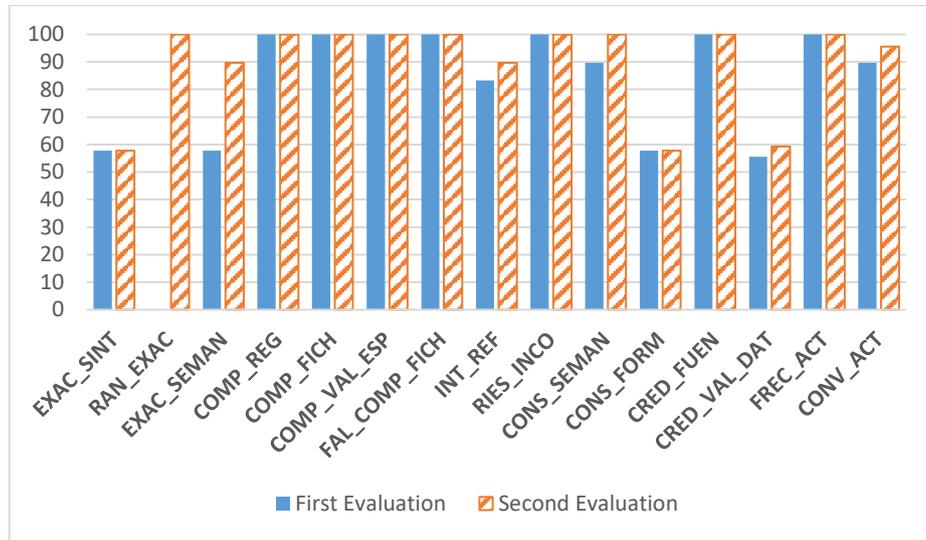

**Fig. 11.** Comparison of results for properties between the first and second data quality evaluation for Org. 3.

## 4. Discussion

In this section we discuss the main findings and lessons learnt that we observed after a deep analysis on the three industrial experiences of application of the evaluation environment. We present these findings from two different point of views that complement each other: the point of view of the evaluation team and the point of view of the organizations that underwent the evaluation. It is important to remark that the findings from the evaluation team were identified after analysing how the work that was carried out during the execution of the data quality evaluation process for the three organizations; on the other hand, the finding from the point of view of the organizations were identified after analysing their feedback and conclusions provided during the activity 5 of the process (Conclude the data quality evaluation), as well as during the posterior follow-up meetings.

The findings from the evaluation team were mainly related to the technological, communicational and organizational concerns when conducting the evaluation process. On the other hand, the findings from the organizations were more related to the obstacles they found in the evaluation process and the benefits that they were provided with by getting insight on the actual levels of quality of their data repositories and the risks regarding their data not meeting their quality expectations and requirements.

### 4.1. Findings from the evaluation team side

As part of the evaluation team, we found that during the application of the data quality evaluation process in the three industrial experiences, some concerns were systematically raised. Following are these findings, grouped by process activity and identified using the acronym FET (Finding from the Evaluation Team).

During the execution of the **Activity 1. Establish the data quality evaluation requirements**, we found the following concerns:

- **FET.01.** The staff of the organizations under evaluation are not usually familiar enough with the concept of "business rules", and its importance when it comes to manage data or to determine data quality. We also noticed that the staff from the organizations found it difficult to identify and describe the business rules that applied to the data they worked with. Due to this reason, a large amount of time was required to develop a first version of the document containing all the business rules that would be evaluated for each data quality characteristic. We found that providing the organizations with a template and some examples of business rules really helps to systematically guide them in the process of elicitation of the business rules and data quality requirements.

- **FET.02.** Even when some of the personnel of the organizations under evaluation are aware of the need for data quality, it is only to a limited extent: they only have the perception or the intuition that the data are of the sufficient quality or not, and usually this perception of data quality is limited to the *"Completeness"* and *"Currentness"* of the data. Other than that, usually no efforts have been formally addressed by the organizations to institutionalize data quality management best practices. In this sense, the most mature organizations tend to widely use reference data as a means to avoid data quality problems, which to some extent, leads to the increase in the quality level of the data.

- **FET.03.** When discussing with the organizations on how to systematize the data quality management initiatives, only a few of them knew the standards that are used at the core of the data quality evaluation process. In this sense, most of the organizations finally realized that both standards, ISO/IEC 25012 and ISO/IEC 25024, could be a kind of data quality evaluation body of knowledge.

- **FET.04.** Many organizational information systems have experienced *ad hoc* maintenance through the years to accommodate some data quality requirements (e.g., business rules for the data). This meant that the business rules became embedded within the newer versions of the information systems, which made it very difficult to keep their traceability. However, it was still possible to extract some information as regards the business rules by inspecting not only the documentation related to the creation of the data repositories, but also the documentation generated after the successive maintenance of the information systems. In addition, and in the lack of documentation, stakeholders' memory and experience proved to be a good source of information when it came to identify business rules, to further detail them and to validate them.

Secondly, for the **Activity 2. Specify the data quality evaluation**, we identified the following finding:

- **FET.05.** Several iterations had been carried out with the staff of each organization to reach a consensus on the definition of the business rules to be included in the scope of the evaluation. In addition, the definitive description of the business rules could not be achieved until several increments and validations with several stakeholders were performed.

Thirdly, during the execution of the Activity 3. Design the data quality evaluation, we discovered the following findings:

- **FET.06.** Only after understanding the evaluation model and the process in more detail, did the organizations commit fully to the execution of the evaluation plan by allocating the adequate number of human resources. Until this point, the organizations usually did not realize the effort that would require on their part to identify the data quality requirements and business rules.

As regards the **Activity 4. Execute the data quality evaluation**, we identified the following concerns:

- **FET.07.** The execution of *"Syntactic Accuracy"* scripts usually take longer time than the execution of the evaluation scripts for any other data quality property. The reason is that these scripts have to check the syntax for the values of all the attributes and records involved, which requires a larger computational effort.
- **FET.08.** The data quality values, and data quality levels obtained for the quality characteristics *"Completeness"* and *"Currentness"* and their related quality properties are usually the highest after the first evaluation because these data quality characteristics are usually more internalised in the vision of data quality that the organizations have. This is coherent with the finding indicated in Activity 1 (FET.02) regarding the knowledge about data quality already existing in the organizations when the evaluation starts.
- **FET.09.** Organizations tend to define and use extensively default values for many attributes with the intention of assuring adequate levels of data quality in their data repositories regarding the completeness of the data. Unfortunately, during the exploitation of the repositories for operational activities, the use of these default values did not typically contribute to improve the levels of quality of the data as much as intended; the bad news is that this practice is commonly found in many data quality evaluation experiences.
- **FET.10.** It is common to suffer deviations for the schedule of the evaluation plan during the execution of the scripts to obtain the base measures for the data quality properties. The reason for this is that the execution of these scripts usually take longer for a repository clone than for the repository "in production". The differences in the execution environment for both data repositories (e.g., the clone is usually deployed on a not so-optimized technological infrastructure) were found as the main reason for these deviations.
- **FET.11.** After analysing the results of the different evaluation experiences, we can provide a summarized list of the most common weaknesses found in the data repositories evaluated:

- o **FET.11.1.** The most common weakness has to do with data values not complying with specific syntax rules.
- o **FET.11.2.** The second one has to do with the existence of records for which the data values of some attributes do not belong to the expected or allowed ranges.
- o **FET.11.3.** The third one is related to the lack of consistency and compliance of specific formats for attributes with a similar data type throughout the data repository.

Finally, in relation to the **Activity 5. Conclude the data quality evaluation**, we encountered the following finding:

- **FET.12.** Low quality levels (< 3) for the data quality characteristics are symptomatic of the existence of data quality issues. Consequently, the presence of these low values in the results of an evaluation requires effort from the organization to identify and fix the source of the underlying issues. In this sense, the improvement scripts provided to the organizations during this activity largely facilitate the identification of the records that are the cause of these low-quality values due to the non-compliance of the business rules defined. The organizations found these scripts to be very helpful in reducing the effort required for that identification of problematic records.

### 4.2. Findings from the side of the organizations evaluated

In this subsection we are going to introduce the most important findings reported by the organization involved in the three industrial experiences during the conclusion of the evaluation (Activity 5) and in the follow-up meetings held after the data quality evaluation finished. We present these findings using the acronym FO (Finding from the Organization) for their identification.

- **FO.01.** Data quality evaluation and certification helps organizations to assure their sustainability in the long term. In this sense, an executive of Org.1 reported that *"It [the certification] allows us to be more competitive, to anticipate the future more easily and to offer better training to future professionals"*. Analogously, directors of Org. 2 agreed that *"One of the benefits of the data quality evaluation is the possibility of conducting more effective sales actions"*. Finally, the CEO of Org. 3 explained that *"Certification will lead the actions aimed at optimizing data management, resulting in the increased performance of the data centre and an overall improvement in the services provided"*; furthermore, he assured that *"The data quality certification will allow us to present ourselves as a Data-Driven company, aware of how we process information and data, and this will help us to analyse the information in order to better serve our customers"* – in this sense, he expected his company would be leveraged by *"The increment of recognition against competitors, and the increment in customer satisfaction and trust"*.

- **FO.02.** Data quality evaluation and certification helps organizations to get insight on their internal (managerial and technological) business aspects, and the ways of working in the organization. The CTO of Org.1. recognized that *"Nowadays, having a document or a set of documents that specify the data requirements is basic for improving the way we work throughout the organization"*; in addition, this executive also stated that the knowledge contained in these documents makes it possible to *"exchange data between the different business areas in an easy way"*. For executives in Org. 2, one of the best consequences of the improvements made as a result of the evaluation and certification process was the chance to elevate the evaluated data repository to the status of *"Single source of truth"* and that *"we can consider our certified data repository as a strategic asset which will contain all the data that are important for the business, and we must use it as that"*.

- **FO.03.** The knowledge acquired through the data quality evaluation and certification process helps to better support the future data quality management initiatives of the organizations. In this sense, the CIO of Org. 2 declared that *"Knowing the data quality weaknesses was useful for the improvement of data quality KPIs"*. Besides, executives of Org. 3 stated that the evaluation process had been useful to *"Mitigate and manage the risks associated with data across the organization, and ensure regulatory compliance and alignment of the data to standards, laws, and best practice guidelines"*.

### 4.3. Proposal of best practices for the data quality evaluation process

From the previous findings and lessons learnt we can derive a set of best practices related to the activities carried out as part of the data quality evaluation process.

As regards the **Activity 1. Establish the data quality evaluation requirements**, we extracted the following best practices:

- **BP.01.** Systematize the definition of business rules (related to FET.01 in section 4.1, and FO.02 in section 4.2), following a predefined, iterative and incremental process that is supported by templates for the elicitation of business rules. This process must also favour and assure a fluid communication between the staff of the organization and the evaluation team. In the earlier stages of the process, more frequent (e.g., daily) communication between the evaluation team and the staff of the organization is recommended. This will enable the possibility of creating small increments for the business rules document, and receiving frequent feedback from the evaluation team. Once a complete first version of the document is stablished, subsequent versions in which the definition of the rules are refined can follow an agreed delivery schedule, aiming to detect early possible deviations. After several iterations, the final version of the business rules document is obtained.

- **BP.02.** Improve the documentation of the data repositories as business rules are identified (related to FET.01, FET.04, FET.05, and FO.02). The data repository documentation should be updated as the business rules document is being created and refined. This can be facilitated through the use of data dictionaries [22], by employing master and reference data techniques [23] or by following an approach based on metadata.

- **BP.03.** <u>Get the commitment of the entire organization</u> for the data quality evaluation process (related to FET.02 and FET.03, and FO.01). Such commitment and collaboration is required, for example, when creating the business rules document, since different points of view (technical, operative, managerial…) need to be brought together. Since usually business and technical staff have different perspectives regarding data quality, their points of view must be shared, discussed, agreed and unified to better specify the data requirements in the business rules document.
- **BP.04.** <u>Define as soon as possible the scope of the data quality evaluation according to the goals of the organization</u> (related to FET.05, FET.06, and FO.01). In this respect, it is important to emphasize that the data quality characteristics can be evaluated and certified separately, allowing them to choose the desired quality characteristics to be certified or excluding them when the minimum quality level required is not achieved.

In relation to the **Activity 4. Execute the data quality evaluation**, we identified the following best practices:

- **BP.05.** <u>Focus the evaluation and the improvement of data quality on the quality properties</u> so as to make the results easier to understand and guide the improvement actions to be taken (related to FET.05, FET.06, and FO.03). Using this approach, the organization can be provided with fully detailed and specific results at the quality property level. In addition, when weaknesses are detected, it is possible to directly point to the specific data quality property, which makes identifying the necessary improvement actions much easier.
- **BP.06.** <u>Provide adequate hardware infrastructure</u> (related to FET.07 and FET.10) for the servers on which the clone of the data repository will be deployed in order to make the execution of the evaluation scripts more efficient. This best practice allows to reduce the duration of one of the most time-consuming activities in the evaluation process. The execution time of the evaluation scripts depends on the complexity of the business rules, the volume of data, and the set-up of the machine on which the data repository clone under evaluation is deployed. For some evaluation scripts that implement complex business rules that execution time can be quite high (in the range of several hours), and they can even take longer if the connection is not stable or is time constrained.
- **BP.07.** <u>Take into account and implement general improvement actions to mitigate or eliminate common weaknesses</u> (related to the findings FET.07, FET.09, FET.10 and FET. 11) Following are some of the common improvement actions that were identified:
  - o **BP.7.01.** <u>Identify and document the business rules related to the syntax required for attributes</u> (related to FET.11.1). To remediate this weakness, it is necessary to implement controls that automate the checking of these business rules, for example, with the use of regular expressions that are checked to validate the values for the attributes before storing them in the data repository.

- o **BP.7.02.** <u>Identify and document the business rules related to ranges in which the data value must be included</u> (related to FET.11.2). To remediate this weakness, the specific data value ranges for the different attributes in the data repository must be specified and checked before inserting the values in the repository.
- o **BP.7.03.** <u>Identify and document the business rules regarding specific format for attributes and make them consistent across the repository</u> (related to FET.11.3). To remediate this weakness, the specific format for attributes or data types that are similar for different entities of the repository must be defined in a consistent manner and checked before inserting the values in the repository. For example, all the attributes in a represent a date must follow the specific format: YYYY-MM-DD: hh:mm:ss.
- o **BP.7.04.** <u>Make controls redundant across the data life cycle</u> (related to FET.07, FET.09, FET.10, and FET.11). It is important to highlight that the controls needed to address data quality concerns should not be implemented only in the data repository, but also in the software systems that interact with it, even when it could seem unnecessary or redundant. For example, to address the weakness mentioned in BP.7.01, a regular expression control can be implemented in the application collects the data, so that its syntax is validated before it is sent to the repository to store it. By doing this, there will be more assurance that the data will already comply with the specific syntax required when it reaches the data repository.

Regarding the **Activity 5. Conclude the data quality evaluation**, the following best practice was identified from the lessons learnt in the three industrial experiences:

- **BP.08.** <u>Provide the improvement scripts with a good categorization</u> (related to FET.12, FO.02, and FO.3). Categorizing the scripts both by data entity and data quality property greatly facilitates the identification of the records that do not comply with the business rules defined. It is also useful for the organization in terms of prioritizing the improvement actions in order to save or adjust costs and effort.

## 5. Threats to validity

In the following subsections, the threats to the validity [24][25] of the application of the data quality evaluation process in the three industrial experiences presents in this paper are analysed in detail.

### 5.1. Construct Validity

The data quality evaluation environment defined is based on the ISO/IEC 25012 and ISO/IEC 25024 standards, which set a common understanding on data quality through the agreement of prominent experts at an international level. Besides, the data quality model proposed focuses exclusively on the inherent data quality characteristics, so that the quality evaluation of any data repository can be performed regardless of its nature or technological aspects. Therefore, thanks to the alignment to these international standards, conceptual discrepancies that could arise between researchers and practitioners have been avoided.

### 5.2. Internal Validity

The threats that can affect the internal validity of the application of the data quality evaluation process are related to the problems that organizations may have to face when identifying and collecting the set of business rules needed for the evaluation. For example, not understanding the data quality characteristics, not having procedures defined, not having enough documentation to infer the business rules, etc. can lead to a waste of time, a waste of organizational resources, or to a loss of efficiency of the data quality evaluation process.

Therefore, to mitigate this threat, during the initial meetings of the project, the data quality evaluation environment and the concepts related to data quality characteristics and data quality properties are conveniently introduced and explained in detail to the organization. Also, an iterative and incremental approach is followed during the creation and definition of the business rules document in order to mitigate the risks of establishing incorrect or incomplete definitions for the business rules. Using this iterative and incremental approach, the organization advances gradually in the identification of business rules while the accredited laboratory verifies the document and provides constant feedback so as to solve possible doubts or mistakes.

### 5.3. External Validity

The threats to external validity are related to the ability to generalize the usefulness of data quality evaluations for different organizations with different technological environments (for example, relational databases, documental databases, big data-based architectures, or any other database technology).

The different contexts of the organizations (technological infrastructure, business rules, number of data entities, etc.) allowed us to mitigate this threat thanks to the generalization of the data quality evaluation process as well as its technology-agnostic nature. In addition, the data quality evaluation process is also domain-agnostic, which enables the application of the process to organizations in any sector or domain. In fact, we have applied the data quality evaluation process to data repositories owned by different organizations in different sectors, like the banking sector, or public administrations, each having their own conceptual particularities as wells as their different technological infrastructure. However, more applications in different technological contexts and sectors are needed in order to further validate the data evaluation environment.

It is also important to state that the researchers have participated in the implementation of the data evaluation environment and have collaborated in the evaluation and certification process of the three industrial experiences presented in this paper, which could be considered a threat. In this sense, and to mitigate this threat, it would be necessary that some other external researchers could replicate the experiences with the data evaluation environment in different organizations and countries.

### 5.4. Reliability of the data quality evaluation process

AQCLab has been accredited by ENAC, the Spanish accreditation body, and in order to get and retain the accreditation it is regularly audited, checking its independence, the freedom of bias, the freedom of vested interests and the reliability of the application of the process. Also, the evaluation results are supervised regularly during the audits by this external, independent, and conveniently authorized third party, ENAC. This assures that the definition (and the specific instances) of the data quality evaluation process remains coherent over the time, free of bias, and free of vested interests.

## 6. Conclusions

It must be recognized that the potential that analytics brings to organizations in this Big Data era, and the consequent need for data has led to an explosive growth in the interest of organizations for acquiring and processing large amounts of data. Being data at the core of today's organizations, data has been elevated to the status of "organizational asset". And, as it is done with other important assets, organizations must become aware of the importance of its quality over time. This awareness involves the need to systematically evaluate the quality of the data. To better support the data quality evaluation process, frameworks are required. In this paper, we have outlined a data quality evaluation environment based on the international standards ISO/IEC 25012 (which defines the data quality characteristics), ISO/IEC 25024 (which defines the data quality properties and the corresponding information required to adapt their underlying measures), and ISO/IEC 25040 (which defines the structure and the foundations of the evaluation process to be tailored by the evaluators).

After applying the evaluation process to the data repositories of three organizations, we have analysed our experience as the evaluation team, and the conclusions reported by the organizations that were evaluated. It is worthy to highlight the following three benefits recognized by the organizations involved:

- Data quality evaluation and certification helps to assure organization sustainability in the long term.
- Data quality evaluation and certification helps to better know the internal aspects (managerial and technological) of the business and the ways of working of the organization.
- The knowledge acquired through the data quality evaluation and certification process helps to better support the future data quality management initiatives of the organization.

Upon the findings of the experiences, we have derived some best practices aimed to improve the efficiency of the data quality evaluation process, as well as the future satisfaction of other companies that may undergo a data quality evaluation, making the process easier and the results obtained in the data quality evaluation process more useful for them.

Another conclusion that we have reached is that the continuous evaluation of the quality of their data leads organizations to a deeper knowledge of their business processes, and this, in turn, leads to a better performance of the organization.

## Acknowledgements

This research is partially funded by Industrial PhD (Ref.: DIN2018-009705), funded by the Spanish Ministry of Science, Innovation and Universities; GEMA: Generation and Evaluation of Models for Data Quality (Ref.: SBPLY/17/180501/000293), funded by the Department of Education, Culture and Sports of the Junta de Comunidades de Castilla La Mancha, and the Fondo Europeo de Desarrollo Regional FEDER,; DQIoT project (Ref.: INNO-20171086 EUREKA Project No. E!11737), funded by CDTI; ECD project (Ref.: PTQ-16-08504), funded by the "Torres Quevedo" Program of the Spanish Ministry of Economy, Industry and Competitiveness; the ECLIPSE project, (Ref.: RTI2018-094283-B-C31) funded by the Ministry of Science, Innovation and Universities, and the Fondo Europeo de Desarrollo Regional FEDER;

## 7. ANNEX A

This annex explains the meaning of the acronyms for each data quality property featured in **Fig. 6** to **Fig. 11**.

| Acronym | Meaning |
|---|---|
| **COMP** | *"Completeness"* |
| **COMP_FICH** | *"File Completeness"* |
| **COMP_REG** | *"Record Completeness"* |
| **COMP_VAL_ESP** | *"Data Value Completeness"* |
| **CONS** | *"Consistency"* |
| **CONS_FORM** | *"Format Consistency"* |
| **CONS_SEMAN** | *"Semantic Consistency"* |
| **CONV_ACT** | *"Timeliness of Update"* |
| **CRED** | *"Credibility"* |
| **CRED_FUEN** | *"Data Source Credibility"* |
| **CRED_VAL_DAT** | *"Data Values Credibility"* |
| **CURR** | *"Currentness"* |
| **EXAC** | *"Accuracy"* |
| **EXAC_SEMAN** | *"Semantic Accuracy"* |
| **EXAC_SINT** | *"Syntactic Accuracy"* |
| **FAL_COMP_FICH** | *"False Completeness of File"* |
| **FREC_ACT** | *"Update Frequency"* |
| **INT_REF** | *"Referential Integrity"* |
| **RAN_EXAC** | *"Range of Accuracy"* |
| **RIES_INCO** | *"Risk of Inconsistency"* |

**Table 10.** Meaning of the quality property acronyms used in Fig. 6 to Fig. 11.